\documentclass[aps,PRL,twocolumn,groupedaddress,floats]{revtex4}
\usepackage{amssymb}
\usepackage{graphicx}

\begin{document}

\title{Out-of-plane momentum and symmetry dependent superconducting gap in Ba$_{0.6}$K$_{0.4}$Fe$_{2}$As$_2$}

\author{Y. Zhang$^{1}$, L. X. Yang$^{1}$, F. Chen$^{1}$, B. Zhou$^{1}$, X. F. Wang$^{2}$, X. H. Chen$^{2}$, M. Arita$^{3}$, K. Shimada$^{3}$, H. Namatame$^{3}$, M.
Taniguchi$^{3}$,  J. P. Hu$^{4}$, B. P. Xie$^{1}$, D. L. Feng$^{1}$}
\email{dlfeng@fudan.edu.cn}

\affiliation{$^1$ State Key Laboratory of Surface Physics, Department of
Physics, and Advanced Materials Laboratory, Fudan University, Shanghai 200433,
People's Republic of China}

\affiliation{$^2$ Department of Physics, University of science and technology
of China, Hefei, Anhui 230027, People's Republic of China}

\affiliation{$^3$ Hiroshima Synchrotron Radiation Center and
Graduate School of Science, Hiroshima University, Hiroshima
739-8526, Japan.}

\affiliation{$^4$ Department of Physics, Purdue University, West Lafayette, IN
47907, USA}

\date{\today}

\begin{abstract}

The three-dimensional band structure and superconducting gap of
Ba$_{0.6}$K$_{0.4}$Fe$_{2}$As$_2$ are studied with high-resolution
angle-resolved photoemission spectroscopy. In contrast to previous results, we
have identified three hole-like Fermi surfaces near the zone center with
sizable out-of-plane or $k_z$ dispersion.  The superconducting gap on certain
Fermi surface shows significant $k_z$-dependence. Moreover, we found that the
superconducting gap sizes are different at the same Fermi momentum for two
bands with different spatial symmetries (one odd, one even). Our results
further reveal the rich superconducting gap structure in iron pnictides, and
provide a distinct test for theories.
\end{abstract}


\maketitle


The discovery of high-$T_c$ superconductivity in the iron-pnictides ignites
extensive studies on these materials. However, the paring symmetry of the
superconductivity is still not settled. Most of the present theories propose a
$s_{\pm}$ nodeless order parameter that changes sign between the hole and
electron Fermi pockets \cite{theory1,theory2}. However, there are conflicting
experimental evidence for the presence of nodes and nodeless superconducting
gaps \cite{nodeEx,nodelessEx}. Furthermore, multiple gap behavior with the gap
values varying from 2$\Delta/k_BT_c\approx$1.6 to 10 has been reported
\cite{gap1, gap3}. One possible cause of these controversies is the multi-band
nature of iron-based superconductors. In contrast to cuprates, all the five Fe
$3d$ orbitals in iron-based superconductors participate in the low-lying
electronic structure, giving a few hole pockets at the zone center and electron
pockets at the zone corner \cite{Kuroki, DJSingh}. The importance of the Fermi
surface topology and their orbital characters has been pointed out by the
recent theories \cite{DHLee,Bernevig}. It is proposed that the presence of node
is determined by the number of bands which cross the Fermi level, and there are
strong anisotropy and amplitude variation of the superconducting gaps on
different Fermi surfaces. Moreover, various physical properties of the
iron-based superconductors are more three-dimensional (3D) than the cuprates.
For example, the isotropy of the upper critical field has been found in
Ba$_{1-x}$K$_x$Fe$_2$As$_2$ \cite{HQYuan}.

Previous angle-resolved photoemission spectroscopy (ARPES) studies
show isotropic nodeless gaps on all the Fermi surfaces in
Ba$_{1-x}$K$_x$Fe$_2$As$_2$ \cite{Ding, Hasan},
BaFe$_{2-y}$Co$_y$As$_2$ \cite{Ding2}, and
Fe$_{1.03}$Te$_{0.7}$Se$_{0.3}$ \cite{Takahashi}. The most
representative and detailed data to date were taken on the optimally
doped Ba$_{1-x}$K$_x$Fe$_2$As$_2$ samples with a $T_c$ of 38~K.  Two
hole pockets were observed near the zone center, and the
superconducting gap on the inner hole pocket is found to be larger
than that on the outer pocket, which is consistent with the
prediction of theories of $s_{\pm}$ pairing symmetry with a gap
function proportional to $|\cos k_x \cos k_y|$. However, there are
still many issues to be resolved. For example, band calculations
predict three hole Fermi surface sheets at the zone center rather
than two. The relationship between orbital characters and
superconducting gaps is yet to be established. Furthermore, band
calculation suggested that the 3D characters of the electronic
structure are important for the magnetism and superconductivity in
the iron-based superconductors \cite{Ganser}. But due to the limited
photon energies used in previous ARPES studies, the superconducting
gap behavior along the out-of-plane momentum ($k_z$) direction has
not been exposed. The resolutions of these issues are important to
the understanding of the superconducting paring mechanism in
iron-based superconductors.

In this Letter, we have studied the $k_z$ dependence of the
superconducting gap in high quality
Ba$_{0.6}$K$_{0.4}$Fe$_{2}$As$_2$ single crystals with ARPES. We
found that the Fermi surface near the zone center $\Gamma$ actually
contains three hole pockets instead of two as previously reported.
By changing the photon energy, we have revealed the 3D character of
the electronic structure and the superconducting gaps. Significant
$k_z$ dependence of the superconducting gap is discovered on one of
the bands. Particularly, we found that at the same momentum, bands
with different symmetries could exhibit very different gap sizes.
Our results provide a more global picture of the gap in the
material, which would help the construction of microscopic models of
the iron-based superconductors.

\begin{figure}[t!]
\includegraphics[width=7.5cm]{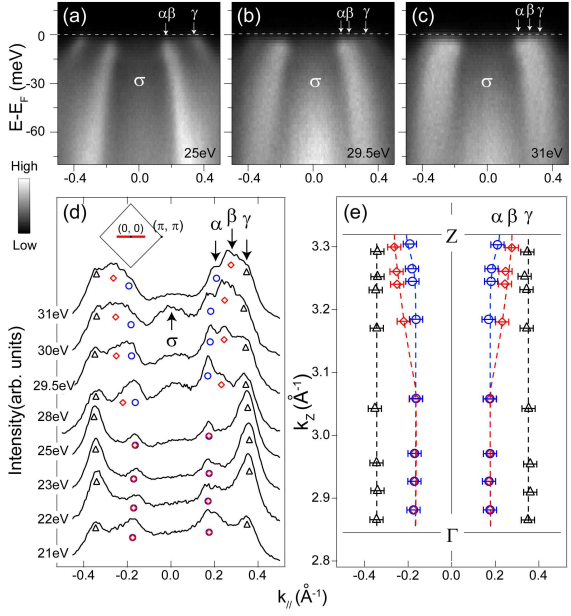}
\caption{(color online) (a), (b), and (c) Photoemission data of
Ba$_{0.6}$K$_{0.4}$Fe$_{2}$As$_2$ along the (0, 0)-($\pi$, $\pi$)
direction taken with 25, 29.5, and 31~eV respectively. (d) The MDCs
at $E_F$ taken with different photon energies are stacked. The inset shows the projection of the cuts in the 2D Brillouin zone. (e) The
$k_F$'s determined from the MDCs in panel d.  Data were taken at 10~K.} \label{Gammaphd}
\end{figure}

High quality Ba$_{0.6}$K$_{0.4}$Fe$_{2}$As$_2$ ($T_c$~=~38~K) single
crystals were synthesized by self-flux method \cite{chen} with a
superconducting transition width of 0.5~K. Data were taken with
various photon energies at the Beamline 5-4 of Stanford Synchrotron
Radiation Laboratory (SSRL), and with 21.2~eV Helium-I$\alpha$ line
of a discharge lamp in mixed polarization geometry. Polarization
dependent data were taken at the Beamline 1 of Hiroshima synchrotron
radiation center (HSRC). Two polarization geometries ($E_p$, $E_s$)
were achieved by rotating the experimental chamber. All the data
were taken with Scienta electron analyzers, the overall energy
resolution is 15~meV at HSRC or 7~meV at SSRL, and angular
resolution is 0.3 degree. The samples were cleaved \textit{in situ},
and measured under ultra-high-vacuum of
$5\times10^{-11}$\textit{torr}.

\begin{figure}[t]
\includegraphics[width=8cm]{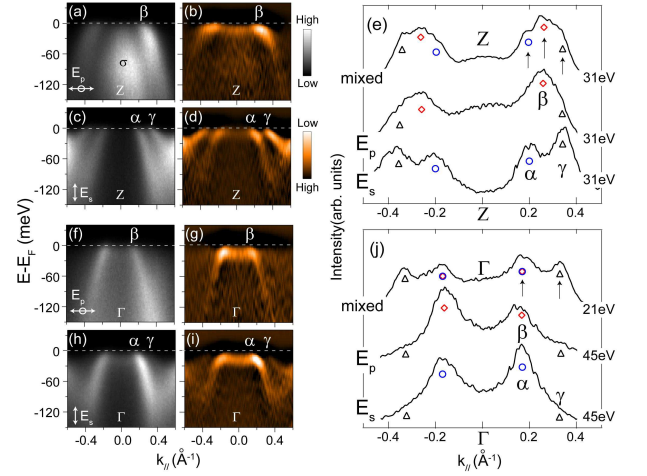}
\caption{(color online)  (a) Photoemission intensity near $Z$ along
the (0, 0)-($\pi$, $\pi$) direction taken in the $E_p$ geometry with
31~eV photons. (b) The second derivative with respect to energy  of
data in panel a . (c) and (d) are the same as panels a and b
respectively but taken in the $E_s$ geometry. (e) The MDCs at $E_F$
with mixed, $E_p$, and $E_s$ polarization geometries near $Z$. (f),
(g), (h), and (i) are the same as panels a, b, c, and d respectively
but taken with 45~eV photons to reach the $\Gamma$ point. (j) The
MDCs at $E_F$ with mixed, $E_p$, and $E_s$ polarization geometries
near $\Gamma$. Data were taken at 40~K.} \label{Gammaorb}
\end{figure}

Since the superconducting gap is almost isotropic around the Fermi surface
cross-section of certain $k_z$ \cite{Ding, Hasan}, we would focus on the $k_z$
dependence of the superconducting gaps along the (0, 0)-($\pi$, $\pi$) high
symmetry cut for simplicity. The photon energy dependent data are shown in
Fig.~\ref{Gammaphd}. Only two bands could be resolved in 25~eV
[Fig.~\ref{Gammaphd}(a)], forming two hole pockets as observed in previous
studies. However, by changing the photon energy, we could observe an additional
bands ($\beta$) moving outwards with 29.5~eV and 31~eV photons
[Fig.~\ref{Gammaphd}(b) and \ref{Gammaphd}(c)]. The Fermi momenta ($k_F$) of
these three bands are determined by peak positions in momentum distribution
curves (MDCs) at the Fermi energy ($E_F$) \cite{FermiCross} in
Fig.~\ref{Gammaphd}(d). Taking the inner potential of 15~eV \cite{Inner} to
calculate the $k_z$'s of different photon energies, our data cover half of the
Brillouin zone along $k_z$ direction from $\Gamma$ with $\sim$~21~eV photons to
$Z$  with $\sim$~31~eV photons [Fig.~\ref{Gammaphd}(e)]. The $\alpha$ and
$\gamma$ bands show little $k_z$ dispersion, while $\beta$  is almost
degenerate with $\alpha$  near $\Gamma$, but moves outward significantly near
$Z$. Since previous ARPES studies were restricted around the $\Gamma$ region,
$\alpha$ and $\beta$ could not be distinguished. Therefore, three bands
reported here with distinct $k_z$ dispersions naturally resolve the previous
inconsistency with band structure calculations, which predicted three hole
pockets around the zone center \cite{DHLee, Bernevig}.

To further reveal the orbital characters of these three bands, we
have conducted polarization dependent ARPES experiment. The
polarization geometry $E_p$($E_s$) could be achieved, with the
polarization direction parallel (perpendicular) to the mirror plane
defined by the sample normal and the (0, 0)-($\pi$, $\pi$)
direction. The orbitals of even (odd) symmetry with respect to the
the mirror plane could be observed in $E_p$($E_s$) geometry
\cite{Zhang09B}. The polarization dependent data around $Z$ are
shown in Figs. \ref{Gammaorb}(a)-\ref{Gammaorb}(d). $\beta$ could
only be observed in $E_p$ geometry, thus it has even orbital
character. Oppositely, $\alpha$ only shows up in $E_s$ geometry,
which suggests an odd orbital character. $\gamma$ mostly has odd
orbital character [Figs. \ref{Gammaorb}(c) and \ref{Gammaorb}(d)],
but there is also some trace of $\gamma$ in $E_p$ geometry due to
possible orbital mixing. The MDCs at $E_F$ of different polarization
geometries around $Z$ are compared in Fig. \ref{Gammaorb}(e). The
peak positions in $E_p$ and $E_s$ geometry are consistent with the
data taken in mixed polarization geometry. Since the different
symmetries of $\alpha$ and $\beta$ should not change along $k_z$
\cite{Ganser}, it enables us to separate them with polarization
dependent experiment around $\Gamma$, as shown in Fig.
\ref{Gammaorb}(f)-\ref{Gammaorb}(i). The MDC peaks of $\alpha$ and
$\beta$ show up at the same momentum in Fig. \ref{Gammaorb}(j). This
agrees well with our photon energy dependent data that the $k_F$'s
of the $\alpha$ and $\beta$ bands are almost degenerate around
$\Gamma$, but separated from each other around $Z$. Note that the
intensity of $\gamma$ is very weak in the polarization data around
$\Gamma$ due to possible matrix element effects of the particular
photon energy and/or experimental setup.

\begin{figure}[t]
\includegraphics[width=8cm]{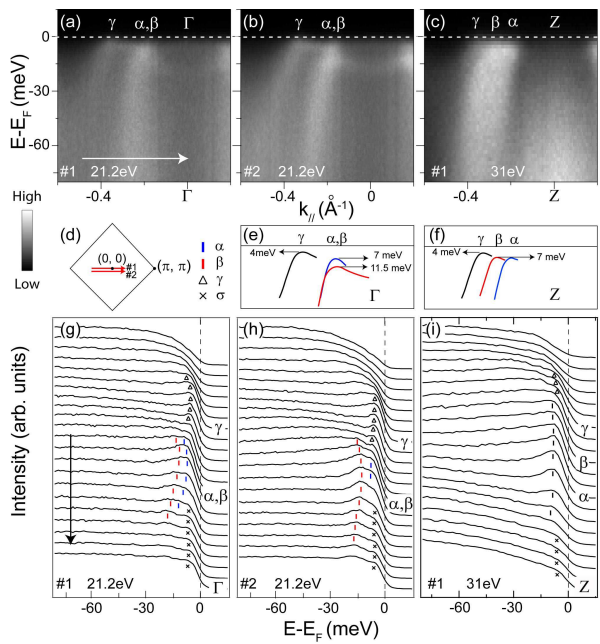}
\caption{(color online) (a) and (b) Photoemission intensities taken
with 21.2~eV along two momentum cuts as indicated in panel d. (c) is
the same as panel a, but taken near $Z$ with 31~eV photon energy.
(e) and (f) are cartoons of the bands in panels a and c
respectively. (g), (h), and (i) EDCs of the data in panels a, b, and
c respectively. Data were taken at 10K.} \label{GammaSC}
\end{figure}

The two odd and one even bands observed here qualitatively agree
with the prediction of the band calculations, where only three
orbitals $d_{xz}$, $d_{yz}$, and $d_{xy}$ contribute to the low
energy electronic structure around $\Gamma$ \cite{DHLee, Bernevig}.
The $\beta$ band is $d_{xz}$ orbital due to its even symmetry. The
$\gamma$ band shows little dispersion along the $k_z$ direction,
which is most likely the $d_{xy}$ orbital with two-dimensional (2D)
character. The $\alpha$ band is thus assigned to the $d_{yz}$
orbital, which is predicted to be degenerate with the $d_{xz}$
orbital at $\Gamma$  by theory. Note that there are some broad
spectral weight ($\sigma$) at (0, 0), which is possibly due to the
contribution of $d_{z^2}$ bands below $E_F$ or the incoherent
spectral weight scattered from other states [Figs.~\ref{Gammaphd}
and \ref{Gammaorb}(a)]; they are thus ignored in the following
discussions.

The data in the superconducting state near $\Gamma$ are shown in
Figs.~\ref{GammaSC}(a) and \ref{GammaSC}(b). The $\gamma$ band shows simple
Bogoliubov dispersion with an energy gap about 4~meV. Most notably, at the
$k_F$'s of the $\alpha$ and $\beta$ bands, the EDCs exhibit a complex structure
with two peaks [Figs.~\ref{GammaSC}(g) and \ref{GammaSC}(h)]. The peak
positions could be tracked towards $\Gamma$ with Bogoliubov like dispersion of
two energy scales (7~meV and 11.5~meV) as shown in Fig.~\ref{GammaSC}(e). The
energy scale of 11.5 meV is consistent with the gap size observed in previous
ARPES experiments \cite{Ding}. Moreover, the clear bending-over behavior and
small peak width clearly reproduce the properties of a superconducting peak.
Furthermore, the peak at 7~meV should not be the bent-over feature of the
$\gamma$ band, as one can track the $\gamma$ band by the triangles in
Fig.~\ref{GammaSC}(g) and \ref{GammaSC}(h). It quickly bends over to high
binding energies and loses its weight significantly. Therefore, the most
natural explanation of our results is that the $\alpha$ and $\beta$ bands have
different superconducting gaps at the same momentum. On the other hand, while
these three bands could be clearly separated around the Z point in
Fig.~\ref{GammaSC}(c), only single peak could be observed at the Fermi
crossings of  $\alpha$ and $\beta$  with a superconducting gap of 7~meV
[Fig.~\ref{GammaSC}(i)].

In the five-band model, the Fermi surfaces of iron-based superconductors
consist of sections with different orbital characters \cite{Kuroki, Bernevig}.
It has been proposed that the multi-orbital interactions could form strong gap
anisotropy on the Fermi surfaces. Therefore, the observation of two energy
scales of $\alpha$ and $\beta$ could be directly related to their orbital
nature, since they show opposite orbital symmetries at the same Fermi momentum
position. That is, the paring strengths could be strongly determined by the
orbital symmetries. The next question is which band provides a larger gap
around $\Gamma$. Considering  there is a significant change from 11.5 meV to 7
meV for one of the gaps along the $k_z$ direction, plus the strong $k_z$
dependence of $\beta$, it is reasonable to assume that the $\beta$ band
contributes to the larger gap of 11.5~meV around  $\Gamma$. We could not
completely exclude the possibility of the $\alpha$ band at this stage, but this
assumption would only affect the gap assignment around $\Gamma$, and would not
affect the observed $k_z$ dependence of the superconducting gaps discussed
below. Polarization dependent experiments with high resolution in the
superconducting state are needed to clarify this point. We leave this for
further studies.

\begin{figure}[t]
\includegraphics[width=7.5cm]{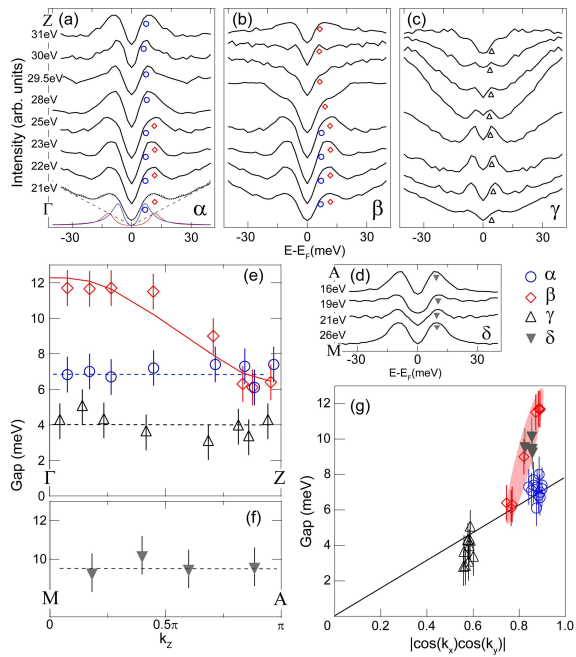}
\caption{(a), (b), and (c) Symmetrized EDCs at the $k_F$'s of three
bands taken with different photon energies. The dashed line in panel
a is the background used in the fitting process. (d) Symmetrized
EDCs at the momentum 0.2~{\AA}$^{-1}$ from ($\pi$, $\pi$), which is
the $k_F$ of electron pocket ($\delta$) at the zone corner
\cite{Ding}. (e) and (f) $k_z$ dependence of the superconducting
gaps around the zone center and corner respectively. (g) Gaps vs.
$|\cos k_x \cos k_y|$.}
 \label{Sim}
\end{figure}

The symmetrized EDCs at the $k_F$'s of the $\alpha$, $\beta$, and
$\gamma$ bands are summarized in Figs.~\ref{Sim}(a), \ref{Sim}(b),
and \ref{Sim}(c). The two-peak behavior could be only resolved
around the $\Gamma$ region. The superconducting gap values are
determined by fitting the peak positions in symmetrized EDCs with
the common phenomenological superconducting spectral function as
illustrated in Fig.~\ref{Sim}(a) \cite{SymEDC}. The $k_z$ dependence
of superconducting gaps is shown in Fig.~\ref{Sim}(e). The gap sizes
of $\alpha$ and $\gamma$ are always about 7 and 4~meV respectively,
and the gap of $\beta$ shows strong $k_z$ dependence  from 11.5 to
7~meV. In previous ARPES studies, these gaps around $\Gamma$ can fit
well to $\Delta_0|\cos k_x\cos k_y|$
\cite{Ding,Hasan,Ding2,Takahashi}. Therefore the superconducting gap
should decrease away from the (0, 0) point. We plot the
superconducting gap versus $|\cos k_x \cos k_y|$ in
Fig.~\ref{Sim}(g). The $\alpha$ and $\gamma$ bands could be well
fitted in this relation   with $\Delta_0$ $\approx$ 8~meV as shown
by the straight line in Fig.~\ref{Sim}(g). However, the large
deviation of the $\beta$ band (highlighted by the shaded region)
indicates that the $k_z$ dependence of the superconducting gap could
not be explained by the in-plane Fermi surface size change of the
$\beta$ band. Therefore, the $k_z$ contribution must be included to
describe the gap function of the $\beta$ band. With the simple
formula of $\Delta_{0\beta}|\cos k_x\cos k_y|(1+A\cos k_z)$, one can
fit the gap of $\beta$ reasonably well  with $\Delta_{0\beta}\approx
11.2~meV $ and $A \approx 0.24$ as shown  by the solid line in
Fig~\ref{Sim}(e).

The  $k_z$-dependent superconducting gaps discovered here could not be
explained in 2D models, which are used to construct the paring mechanism in
most of previous theories \cite{DHLee, Bernevig}. In contrast to the 2D
electronic structure of cuprates, the 3D characters of iron-based
superconductors have been highlighted by many properties, \textit{e.g.}, the
small anisotropy of the upper critical field \cite{HQYuan, HQYuan2}, and 3D
spin fluctuations in parent and doped compounds \cite{Zhaoj, INS1, INS2}. Our
results further emphasize the 3D character of the superconductivity, where the
out-of-plane paring channels should be considered. Consistently, a substantial
$k_z$ dependence of the superconducting order parameter has been found in a
recent 3D band model that is constructed to calculate the spin fluctuations and
the paring function \cite{Ganser}.

The superconducting gap size is proposed to relate to the nesting
condition of Fermi surfaces between $\Gamma$ and $M$ in previous
ARPES studies \cite{Ding, Ding2}. The Fermi surfaces near the zone
corner show weak $k_z$ dependence\cite {M3D1, M3D2}. In that case,
the change of Fermi surface size of $\beta$ along $k_z$ direction
could break the nesting condition, and thus significantly change the
superconducting gaps. However, it could not explain the different
gaps of the $\alpha$ and $\beta$ bands, since they have almost the
same Fermi surface size around $\Gamma$. Moreover, the $k_z$
dependence of gaps near the zone corner are not so obvious, as shown
in Figs. \ref{Sim}(d) and \ref{Sim}(f), and the gap sizes also
deviate from the fitting of $|\cos k_x \cos k_y|$ relation in Figs.
\ref{Sim}(g). Therefore, the paring between $\Gamma$ and $M$ could
not be just related to simple nesting of the Fermi surfaces. The
orbital character and the 3D electronic structure should be taken
into account.

To summarize, we have carried out a systematic investigation of the
superconducting gap of high quality
Ba$_{0.6}$K$_{0.4}$Fe$_{2}$As$_2$ single crystals, and have
established a direct connection between the superconductivity and
the 3D electronic structure with multi-orbital nature in iron-based
superconductors. Our results have set up a more comprehensive
picture of the superconducting gap in this compound, which shed new
light on the understanding of superconductivity in iron-based
superconductors.

We gratefully acknowledge the experimental support by Dr. D. H. Lu
and Dr. R.G. Mooreat at SSRL. This work was supported by the NSFC,
MOE, MOST (National Basic Research Program No.2006CB921300), and
STCSM of China. SSRL is operated by the US DOE Office of Basic
Energy Science.

\end{document}